\documentstyle{mn}
\input{epsf}
\def\n{{\noindent}}

\title[Evolution of correlation functions]
{Modelling the evolution of correlation functions in
gravitational clustering \\} 

\author[Munshi $\&$ Padmanabhan]{Dipak Munshi$^1$, 
T. Padmanabhan$^2$\\
$^1$Queen Mary and Westfield College, London E1 4NS, United Kingdom \\
$^2$Inter-University Centre for Astronomy and Astrophysics,
Post Bag 4, Ganeshkhind, Pune 411 007, India\\
\smallskip
Email: D.Munshi@qmw.ac.uk, nabhan@iucaa.ernet.in\\
}

\pagerange{000,000}

\begin{document}

\maketitle

\begin{abstract}
Padmanabhan (1996) has suggested a model to  relate the nonlinear two - point correlation function
to the linear two - point correlation function. In this paper, we extend this model in two directions: (1) By averaging over the initial Gaussian distribution
of density contrasts, we estimate the spectral dependence of the scaling
between nonlinear and linear correlation functions. (2) By using a physically
motivated ansatz, we generalise the model to N-point correlation functions
and
relate the nonlinear, 
volume averaged, N-point correlation function $\bar\xi_N(x,a)$ with linearly extrapolated volume averaged 2-point correlation function $\bar \xi_2(l,a)$ evaluated at a different scale.  We compare the point of transition between different regimes obtained from our model with numerical simulations
and show that the spectral dependece of the scaling relations seen in the
simulations can be easily understood. Comparison of the calculated form of
 $\bar\xi_N$ with the simulations show reasonable agreement. We discuss several
implications of the results.
\end{abstract}


\section{Introduction}

There is growing evidence that the large scale structure in the universe formed through gravitational amplification of small inhomogeneities. 
Semianalytic modelling
of gravitational clustering of collisionless, non relativistic, dark matter 
particles will be of significant utility in understanding the
formation of large scale structures. Such a modelling is straightforward
when the density contrasts are small and perturbation theory, based on a
suitably chosen small parameter, is applicable (see e.g., Fry 1984, Moutarde et al 1991, Buchert 1992, Bernardeau 1992 ).
In the other extreme, highly nonlinear regimes can be handled if one is
prepared to make some extra assumptions like stable clustering ( Peebles 1980 )
or those which underly the Press-
Schecter formalism, Peaks formalism etc. ( Bardeen et al. 1986 ). The intermediate
regime is considerably more difficult but some progress has been made 
recently even in this case  (  Padmanabhan, 1996; also see Hamilton et al. 1991,
Nityananda and Padmanahban 1994), using the scale invariant spherical infall models. These papers give an expression for the nonlinear mean correlation function in terms of the linear mean correlation function both in the intermediate and nonlinear regimes. To do so, 
Padmanabhan (1996) has concentrated on a typical spherical region  and has ignored the effects arising out of averaging over peaks of different sizes. Also no attempt was made to model higher order correlation functions.      
In this paper, we generalise these ideas in two directions: 

(a) We consider peaks of different heights and average over them along the lines
suggested in Padmanabhan et al. (1996). Such an averaging will introduce
spectrum dependence in the relation connecting nonlinear and linear correlation
functions which have been noted in numerical simulations ( Padmanabhan et al.
1996, Jain et al. 1995, Peacock \& Dodds 1996). We will show that this dependence can be
understood from our model. 

(b) We shall postulate an ansatz for the higher
order correlation functions, generalising the result for two point correlation
function. Using this ansatz, we shall calculate the $S_N$ parameters for
both the intermediate and nonlinear regimes and compare them with the
simulations. We will see that there is reasonable - though not excellent -
agreement between the model and simulations suggesting that the ansatz we have
proposed is along the right direction.

\section{The model and the ansatz}

The basic idea behind the model used in Padmanabhan ( 1996 ) can be described
as follows: Consider the evolution of density perturbations starting from
an initial configuration which we take to be a realisation of a Gaussian
random field with variance $\sigma$. A region with initial density contrast $\delta_i$ will expand
to a maximum radius $x_{ta} = x_i/ \delta_i$ and will finally collapse to an object of radius $x_f$ which will contribute to the 
two-point correlation function an amount proportional to $(x_i/x_f)^3$. The initial density contrast within a 
{\it randomly} placed
sphere of radius $x_i$ will be $ \nu \sigma (x_i)$ with a probability
proportional to $\exp (-\nu^2/2)$. On the other hand, the initial density 
contrast within a sphere of radius $x_i$, {\it centered around a peak in the 
density field} will be proportional to the two-point correlation function 
and will be  $\nu^2 \bar\xi (x_i)$ with a probability proportional to $\exp (-\nu^2/2)$. [We have obtained the quadratic scaling in $\nu$ based on
the assumption that $\bar\xi$ scales in the same way as mean square fluctuations
in the mass, which - in turn - will scale as the mean square of the gaussian
density field. In general, one expects the scaling to be $\nu^{\alpha}$ with
$\alpha\approx 2 - 1$. The results are easily generalisable to any value of
$\alpha$. We will stick to $\alpha =2$ since it gives reasonable agreement
with simulations and is based on simple considerations ]. It follows that the contribution from a {\it typical} region will 
scale as  $ \bar \xi_{nl} 
\propto \bar \xi_i^{3/2}$ while  that from {\it higher peaks} will scale as $\bar \xi_{nl} 
\propto \bar \xi_i^3$. In the intermediate phase, most dominant contribution
arises from high peaks and we find the scaling to be  $\bar \xi_{nl} 
\propto \bar \xi_i^3$. The non-linear virialized regime is dominated by 
contribution from several typical initial regions and has the scaling
 $\bar \xi_{nl} 
\propto \bar \xi_i^{3/2}$. This was essentially the feature pointed out in 
Padmanabhan (1996) though in that work it was assumed that $\nu = 1$. 
To take into account the statistical fluctuations of the initial Gaussian
field we can average over different $\nu$ with a Gaussian probability 
distribution. [Strictly speaking, there will be deviations from pure
gaussian distribution because our averaging requires a mapping from
lagrangian to eulerian coordinates; we shall ignore this because it is
a higher order effect]. We shall do this calculation in the next section.

To generalise the above ideas for higher order correlation functions is
more nontrivial. 
In general n-point correlation functions will depend on shapes but 
volume averaging will remove this shape dependence. The  ``$S_N$ parameters"  are
then defined as dimensionless ratios of  $\bar \xi_N(x,a)$ and $\bar \xi_2(x,a)^{(N-1)}$.  Such
volume-averaged N-point functions 
( which can  be 
directly related to counts-in-cells ) and the $S_N$
parameters have been studied extensively
in literature ( White 1979, Balian \& Schaeffer 1984, 
Bouchet et al. 1991, Bouchet \& Hernquist 1992)
 . The $S_N $ parameters
show fairly simple pattern of behaviour both in the perturbative
and nonlinear regimes.
It can be shown that all $S_N$'s  can be evaluated from spherical 
collapse model in the limit $\xi_2 \rightarrow 0$. In this limit, 
they are constant and depend only on initial  spectral index when 
smoothing is taken into account. They are also
expected to be constants in the nonlinear regime. These results indicate that 
the hierarchical pattern, which is generally assumed to describe nonlinear 
$\xi_N$ functions, could have a larger range of validity. We shall exploit this possibility to
estimate the $S_N$ in the intermediate and nonlinear regimes along the
following lines:

 The evolution of N-point correlation functions is
described by momentum moments of BBGKY hierarchical equations, which can
be expressed in the form

\begin{equation}
{\partial Q_N \over \partial t} + {1 \over a} \sum { \partial \over 
\partial x_i^{\alpha}} (Q_N v_i^{\alpha}) = 0.  
\end{equation}
Here $\alpha$ varies from 1 to N,  $i$ varies over the Cartesian components and $Q_N$ is the full N-point correlation function given by

\begin{eqnarray}
Q_N(1,2,...N) = 
1 &+& \xi_2(1,2)  + ....  \nonumber  \\
+~~ \xi_3(1,2,3) &+& .... 
+~~ \xi_N(1,2,3,...N) 
\end{eqnarray}
and $\xi_N$ denotes the reduced part of N-point correlation function.
For 2-point correlation function, the resulting 
equation can be simplified to ( Peebles 1980 )

\begin{equation}
{\partial \xi_2 \over \partial t} + {1 \over ax^2} {\partial \over \partial x}
\big( x^2 ( 1 + \xi_2)v \big ) = 0.
\label{three}
\end{equation}
which describes the conservation of pairs. In the integral form, the same result
can be expressed as 
\begin{equation}
x^3( 1 + \bar\xi_2(x)) = l^3  
\end{equation}
where $l = 
\langle x_i^3 \rangle  ^{1/3}$
 is the average initial scale
from which collapsed structures of size $x$ have formed. 

Our aim is to generalise the above result for higher order correlation
functions, but it is obvious that one can not get such a simple 
relation for higher order correlation functions in which
$N-1$ different length scales are present. 
To make progress, one needs to assume that,  although there are 
different length scales present in reduced n-point correlation function,
all of them have to be roughly of the same order to give a significant contribution. This is
supported by the fact that - by its very construction - the reduced N-point 
correlation function vanishes when a single point or group of points 
from this set of $N$ points are moved to large separation (In that 
limit, the correlation function is just the product of lower order
reduced correlation functions). For a geometrical picture, one can think of a  polyhedron, inscribed in sphere, with particles
at each vertex having their velocities directed towards the centre of the sphere. (This configuration has the relative 
velocity of particles directed along their  relative separation
and hence can satisfy the stable clustering ansatz.) For such a configuration,
the scale in the correlation function will be the radius of the sphere circumscribing the polyhedron. If the correlation functions are described
by a single scale, then a natural generalisation of equation (4), will
be

\begin{equation}
\bar \xi_N \approx \langle x_i^{3(N-1)} \rangle
/ x^{3(N-1)}
\label {xibarn}
\end{equation}

The validity of such an ansatz is 
open to question and we do hope to check it directly in numerical
simulations in a future work.
In this paper we accept the above ansatz as a working hypothesis
and use it to calculate the $S_N$ parameters 
in different regimes. Since these parameters have been studied numerically we can directly test  
predictions of this ansatz against existing results to obtain a feel for
the validity of the ansatz. 
It may be noted that, even though several models has been proposed to predict
the values of $S_N$ parameters ( Hamilton 1988, Fry 1984, Schaeffer 1984, Balian \& Schaeffer 1989, Bernardeau \& Schaeffer 1992 ) they 
fail to reproduce correct values for lower order $S_N$ parameters.
We shall 
show that the non linear values of lower order $S_N$ parameters are 
predicted with fair degree of accuracy in our model. 

\section{The correlation functions in different
regimes}

We shall now consider the implementation of the above ideas in three
different regimes of gravitational clustering. We shall call the first
one `` perturbative regime " in which we expect perturbation theory is valid.
The second regime ( which we call `` intermediate " regime ) is dominated by
scale invariant radial infall of high peaks. Finally the third regime (`` nonlinear ") is dominated by virialised blobs of matter. While we are mainly interested in the latter two regimes, we shall begin with some important
observations regarding the perturbative regime.

\medskip\noindent
{\it (a) Perturbative Regime}
\smallskip\noindent

We divide the density field into two  parts at each point with 
one part coming from spherical collapse 
( which we call the  `` monopole " part ) and the rest of the contribution comes
from higher order spherical harmonics, characterising shear, tide and nonlinear
coupling between them: 
\begin{equation}
\delta(x,a) = \delta_{sph}(x,a) +  \epsilon(x,a)
\end{equation}
It should be noted that one can {\it not} assume $ \delta(x) = \delta_{sph}(x)$
for each point. While this may be obvious from symmetry considerations, a more
formal argument can be given along the following lines: Let us assume
for a moment that we can set $\epsilon=0$. Since the growth of density contrast in spherical collapse model
is well known (Peebles 1980), we can expand $\delta_{sph}(x,a)$ in taylor series 
to get 

\begin{equation}
\delta(x,a) = \delta_{sph}(x,a) = \sum^{\infty}_{N=1} \delta^{(N)}_{sph}(x,a) = \sum^{\infty}_{N=1}{\mu_N \over N!}\delta^{(1)N} 
\end{equation}
where $\mu_2 = 34/21, \mu_3 = 682/189$, $\mu_4 = 446,440/43,659$....
It is clear that, in spherical collapse, $\delta^{(N)} \propto \delta^N$. 
On the other hand, since 
$\langle \delta \rangle = 0$  we demand $\langle \delta^{(N)}
\rangle = 0$ at every order of perturbation. This implies, in case of spherical collapse model, that 
$\langle \delta^N \rangle = 0$ for all N i.e., moments of $\delta^N$ vanish
at all orders; hence we must conclude that $\delta$ vanishes at each point identically. Clearly, we cannot set $\epsilon=0$ in (6).

For a generic Gaussian field, we have to work with a $\delta$ which has 
two parts, one coming from pure spherical collapse the other part, $\epsilon$
is related to deviation from spherical collapse dynamics. The taylor series
will then be
 
\begin{equation}
\delta(x,a) = \sum^{\infty}_{N=1} \delta^{(N)}(x,a) = \sum^{\infty}_{N=1}{\mu_N \over N!}\delta^{(1)N} + \sum^{\infty}_{N=2}
\epsilon_Na^N 
\end{equation}
where we have expanded  $\epsilon$ in a perturbative series with
$\epsilon_N$ being  of order $\delta^N$. ( Note that 
there is no contribution to $\epsilon$ in linear order ). Although  both the terms will be important for a generic random field, it has already been shown ( Bernardeau 1994a) that $\epsilon$ becomes less dominating
for rare events i.e. for large values of $\nu$ =($\delta / \sigma$). In 
the perturbative regime, where $\sigma$ is small,  any deviation from homogeneity
is a rare event and hence one can assume 
that, statistically, $\epsilon$ will be close to zero at most of the points.
One can explicitly demonstrate this claim by calculating the parameters $S_N
 = \langle \delta^N \rangle_c/\langle 
\delta^2 \rangle_c^{(N-1)}$  in the limit  $\epsilon_N \rightarrow 0$
and showing that it will reproduce the well known results of $S_N$ derived earlier by 
summing up all tree level diagrams in the limit $\sigma \rightarrow 0$ (Bernardeau 1992 ).
Consider, for example, the case of $S_3$; we have

\begin{eqnarray}
S_3 = \langle \delta^3 \rangle / \langle \delta^2 \rangle^2 &=&
3\langle {\delta^{(1)}}^2\delta^{(2)}\rangle/\langle {\delta^{(1)}}^2\rangle^2\\
&=& 3(\langle {\delta^{(1)}}^2 \rangle \langle 1/2\mu_2 {\delta^{(1)}}^2 + \epsilon_2
\rangle \nonumber \\
&+& \langle \epsilon_2 \delta^{(1)^2} \rangle + \mu_2 \langle {\delta^{(1)}}^2 \rangle^2)/ \langle {\delta^{(1)}}^2 \rangle^2)
\end{eqnarray}
The first term vanishes because $\langle \delta^{(2)} \rangle = 0$ and the 
second term gives us vanishing contribution in the limit $\epsilon_2 \rightarrow 0$
and we get the well known result $S_3 = 3\mu_2$.
A similar calculation for higher order moments reproduce tree level
results of perturbation theory, $S_4 = 4\mu_3 + 12\mu_2^2$ etc.
These are the exact values of $S_N$ parameters in the limit $\sigma^2 \rightarrow  0$ (i.e. at the tree level of perturbative calculation neglecting
all loop corrections ) obtained previously by rigorous analytical 
calculations 
( Bernardeau 1992, 1994a, 1994b, 1994c, 1995 ). Our analysis reconfirms that any deviation from spherical dynamics does not alter the values of $S_N$ parameters at tree level  in which 
only monopole part of the dynamics is relevant. The higher order
harmonics ( shear, stress and their couplings ) start contributing
only from loop level. ( This is also true for 
approximation schemes like Zeldovich approximation etc; see Munshi et al. 1994 )

\bigskip\noindent
{\it (b) Intermediate regime}
\medskip\noindent
 
In the intermediate regime, we concentrate on the collapse of regions
around peaks in which the density contrast scales as the correlation
function. [We shall work with a $\Omega =1 $ universe]. 
Consider a spherical region of initial radius $x_i$ and overdensity $\delta
= \nu^2 \bar\xi_L(x_i)=\sigma_0^2 \nu^2 x_i^{-(n_p+3)}$ where $n_p$ is
the index of the initial power spectrum and $\sigma_0$ is a constant. This region will expand to a maximum
radius $x_{ta} = (x_i/\delta) \propto \nu^{-2} x_i^{(n_p + 4)}$ and then collapse
back to a final radius $x_f\propto x_{ta}$. In the scale invariant radial collapse, the resulting profile
will scale with $x_{ta}$  ( Fillmore \& Goldreich 1984, Bertschinger 1985, Hoffman \& Shaham 1985). Taking $x=\lambda x_{ta}$ 
and using equation (4), it is easy to see that

\begin{equation}
\bar \xi_2(x) \simeq \big ({\sigma_0\over\lambda^{1/2}} \big )^z  \langle \nu^z \rangle
x^{-3(n_p + 3)/(n_p +4)}
\label {xitwo} 
\end{equation}
where we have introduced the notation $z = 6/(n_p + 4)$.

Evaluating the average $<....>$ using the Gaussian distribution
we find that the final result can be written as $ \bar \xi_2(x) = A \bar \xi_{2,lin}^3(l)$ where
$l\approx x \bar \xi_2^{1/3}$ and 
\begin{equation}
A = {\langle \nu^z \rangle^{6/z} \over \lambda^3}={1\over \lambda^3}\bigg ( {1 \over \sqrt{2 \pi}} 2^{(z -1)/2} \Gamma{((z + 1)/2)} \bigg )^{6/z}
\end{equation}
 The above result is the generalisation [in the intermediate regime] of the analysis presented in
 Padmanabhan ( 1996 ) taking into account the averaging over
different $\nu \sigma$ peaks. It shows that the averaging introduces a
spectrum dependent scaling.

Let us now consider the higher moments.
Using our ansatz for higher order moments (equation (5)) 
we can now compute the result for $\bar \xi_N$ to be 

\begin{eqnarray}
\bar \xi_N(x)= \big ( {\sigma_0\over\lambda^{1/2} }\big )^{z(N-1)}\langle \nu^{z(N-1)}\rangle x^{-3( N-1)(n_p + 3)/(n_p +4)} 
 \end{eqnarray}
The scaling we get for higher order moments is clearly hierarchical in nature.
Using the definition of $S_N$ parameters we find that in this intermediate regime 

\begin{equation}
S_N^{int} = \bar \xi_N /\bar \xi_2^{(N-1)} = \langle \nu^{(N-1)z} \rangle
/\langle \nu^z  \rangle^{(N-1)}
\end{equation} 
or, equivalently,

\begin{equation}
S_N^{int} = (4 \pi)^{(N - 2)/2}{\Gamma \big (  
{z ( N - 1 )  + 1  \over 2 } \big) \over~~~~~~~
\Gamma \big (  {z + 1  \over 2 } \big )^{(N - 1)}}
\end{equation}
Using the above results, we  can also directly relate the $\bar \xi_N(x)$ with $\bar \xi_2(l)$ and otain

\begin{equation}
\bar \xi_N(x) = S_N \bar \xi_2^{(N-1)}(x) = S_N A^{N-1}\bar \xi_{2,lin}^{3(N-1)}(l)
\end{equation}

\bigskip\noindent
{\it (c) Nonlinear Regime}

\medskip\noindent
In this case, we take the initial density contrast to scale as the
variance of the Gaussian random field, so that $\delta\propto x_i^{-(n_p+3)/2}$.
We assume as before that a patch with initial radius $x_i$ will attain a maximum radius $x_{ta} = x/\delta$ which will the collapse to form structure
of size $ x = \lambda x_{ta}$. Then, a corresponding calculation gives 
 
\begin{equation}
\bar \xi_2(x) =  \big ({\sigma_0\over\lambda} \big )^y  \langle \nu^y \rangle  x^{-3(n_p + 3)/(n_p +5)}
\label {xibartwo}
 \end{equation}
where we have introduced the notation $y = 6/(n_p+5)$.
After averaging over the initial Gaussian distribution, this
result becomes $\bar \xi_2(x) = B \bar \xi_{2,lin}(l)^{3/2}$ where
\begin{equation}
B = {\langle \nu^y \rangle^{3/y} \over \lambda^{3}} = {1\over \lambda^{3}}\bigg ( {1 \over \sqrt{2 \pi}} 2^{(y -1)/2} \Gamma{((y + 1)/2)} \bigg )^{6/y}
\end{equation}
This generalises the corresponding result of Padmanabhan (1996) to the
nonlinear regime  by taking into account the initial Gaussian fluctuations. The averaging introduces a spectrum dependent prefactor.

For higher order moments analysis can be done in a equivalent way and 
the result is
\begin{eqnarray}
\bar \xi_N(x) = \big ( {\sigma_0 \over \lambda} \big )^{ y(N-1)} \langle \nu^{y(N-1)}\rangle x^{-3( N-1)(n_p + 4)/(n_p + 5)} 
\end{eqnarray}
where  $y = 6/(n+5)$. So the scaling we get for higher order moments is again hierarchical in nature and the $S_N$ parameters can be evaluated exactly in a same manner as before, giving:
 
\begin{equation}
S_N^{non} = \bar \xi_N /\bar \xi_2^{(N-1)} = \langle \nu^{(N-1)y} \rangle
/\langle \nu^y \rangle^{(N-1)} 
\end{equation} 
or, equivalently

\begin{equation}
S_N^{int} = (4 \pi)^{(N - 2)/2}{\Gamma \big (  
{y ( N - 1 )  + 1  \over 2 } \big) \over~~~~~~~
\Gamma \big (  {y + 1  \over 2 } \big )^{(N - 1)}}
\end{equation}
This result can also be expressed as

\begin{equation}
\bar \xi_N(x) = S_N \bar \xi_2^{(N-1)}(x) = S_N B^{N-1}\bar\xi_{2,lin}^{3(N-1)/2}(l)
\end{equation}

The averaging process $< ... >$ in both quasi-linear and nonlinear regimes
 can be made more
sophisticated by introducing an additional weight factor which is proportional
to some power of Lagrangian volume of the patch from which the object is collapsing i.e. $x_i^m$ (see Padmanabhan et al. 1996). In that case the results generalise to
\begin{equation}
\bar \xi_N(x) = \langle x_i^{3(N-1) + m}\rangle/ 2 \langle x_i^m \rangle x^{3(N -1)} 
\end{equation}
which can be simplified to 
\begin{equation}
S_N = (2\langle \nu^{m \beta} \rangle)^{N-2} \langle \nu^{3(N-1)\beta + m \beta } \rangle/ \langle \nu^{ 3\beta +m\beta} \rangle^{( N - 1)}
\end{equation}
where $ \beta = x/3$ in quasi-linear regime and $y/3$ in nonlinear regime.
Finally the $S_N$ parameters are recovered after doing the averaging as before
\begin{equation}
S_N = \big( 2\Gamma \big( {m\beta + 1 \over 2} \big ) \big)^{ N - 2} {\Gamma \big (  
{3\beta ( N - 1 ) + m\beta  + 1  \over 2 } \big) \over~~~~~~~
\Gamma \big (  {3\beta + m\beta  + 1  \over 2 } \big )^{(N - 1)}}
\end{equation}
The simplest choice is $m=0$ which we shall use in this paper. It may be noted
that the model used by Jain et al. (1995) corresponds to $m=3$; the expressions
given above can be used to read off the $S_N$ parameters in any other scheme.
(As we shall see in the next section, $m=0$ seem to give fairly good fit
to the numerical simulation results ).
We may also note that: 

(i) The expressions derived for $S_N$ parameters are valid for $N\ge 2$
($S_1 = 1$ by definition ).

(ii) Direct comparison with results of intermediate and nonlinear regime shows $S_N^{int}(n_p) = S_N^{non}(n_p + 1)$. Also note that the value of $S_N$ is independent of  $\lambda$.

(iii) Temporal dependance of $\xi_N$ in both quasi-linear and nonlinear regime
can be derived from the fact that any statistic of scale invariant system
can be expressed as a function of $x/x_{nl}$, where $x_{nl}$ is the scale defined through the relation 
$\sigma(x_{nl}) = 1$. Since $x_{nl} \propto a^{2/(n+3)}$ all correlations will
be function of single variable $q = xa^{-(n+3)/2}$. 

(iv) It is clear that except for calculating the averages of powers of $\nu$ ( which
is assumed to be distributed normally ) nowhere have we actually used the fact
that the initial density distribution was Gaussian, which clearly show that
our method of analysis can be generalised in a straight forward manner
to calculate $S_N$ parameters for initially non Gaussian distributions.

(v) For studying gravitational clustering in dimensions other than 3 the same method of analysis can be used with  the scaling 
 $\bar \xi_N(x,a) \propto \bar \xi_{2,lin}^{D(N-1)}(l,a)$
in intermediate regime and $\bar \xi_N(x,a) \propto \bar \xi_{2,lin}^{D(N-1)/2}(l,a)$ in highly nonlinear regime. (vi) Given the $S_N$ parameters, one can compute the void probability
distribution function and related quantities. This calculation is indicated in the Appendix.

\bigskip\noindent
{\it (d) Transition between the regimes}
\medskip\noindent

Having determined the behaviour of correlation functions in the three different regimes, one can enquire where the transition between the regimes occur.
Since there exists three distinct phases in gravitational
clustering we have two transition points: (1) Transition from the perturbative
regime to intermediate regime and (2) Transition from intermediate regime to 
nonlinear regime. Let the first transition occur when $\bar \xi_{2,lin}(l) = T_{c_1}^{(2)}$ and the second when $\bar \xi_{2,lin}(l) = T_{c_2}^{(2)}$. Finding the value of $\bar \xi_{2,lin}(l)$ for which quasi-linear and intermediate
$\xi_2(x)$ matches we get $T_{c_1}^{(2)} = 1/A^{1/2}$. Similarly, equating the
expressions for the intermediate and nonlinear regimes gives  $T_{c_2}^{(2)} = ( B/A )^{2/3}$.

We have used the two point correlation function to define the transitions
since they are most directly related to the density inhomogeneity. It is, of course, possible to repeat the same exercise using higher order correlation
functions. Our results for the higher order correlation functions  can
be summarised as
\begin{eqnarray}
\bar \xi_N^{pert}(x) &=& S_N^{tree} \bar \xi_{2,lin}^{N-1}(l) \nonumber \\
\bar \xi_N^{int}(x) &=& A^{N-1} S_N^{int} \bar \xi_{2,lin}^{3(N-1)} (l) \nonumber \\
\bar \xi_N^{non}(x) &=& B^{N-1} S_N^{non}  \bar \xi_{2,lin}^{3(N-1)/2} (l)  
\end{eqnarray}
Using these, it is easy to see that the transition points defined through
 N-point correlation function will give us
\begin{equation} 
T_{c_1}^{(N)} = ( S_N^{tree} / S_N^{int} )^{1/2(N - 1)} T_{c_1}^{(2)}
\end{equation}
and
\begin{equation}
T_{c_2}^{(N)} = ( S_N^{non} / S_N^{int} )^{2/3(N - 1)} T_{c_2}^{(2)}
\end{equation}
Since $S_N^{tree}< S_N^{int}$ and $S_N^{int} > S_N^{non}$,  it is clear 
that transition for higher order moments will occur for smaller and
smaller values of $\xi_2(l)$.

It should be noted that all though $S_N$ parameters are insensitive
to the  modelling parameters like $\lambda$ the transition points 
are  sensitive to the choice of these variables. We have taken $\lambda = 1/2$
which is close to  value taken  by Jain et al. (1995) for their 
fitting function.

\begin{figure}
\protect\centerline{
\epsfysize = 3.5truein
\epsfbox[20 146 587 714]
{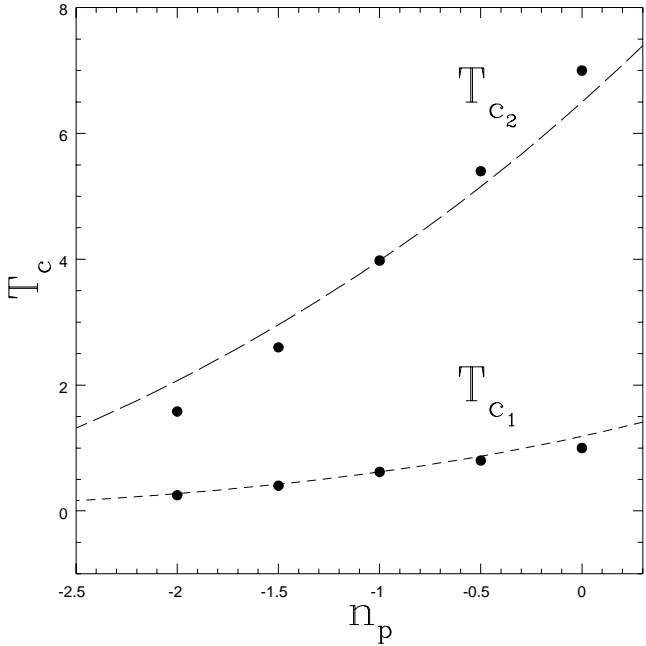}
}
\caption{ Transition points $T_{c_1}$ and $T_{c_2}$ are plotted 
as a function of 
spectral index $n_p$. The short-dash curve corresponds to transition points from perturbative regime to intermediate regime while the long-dash correspond to transition
from intermediate to nonlinear regime, predicted from our model. The curves
are  normalised to
match Jain et al.'s simulation result for $n = -1$. Circles
corresponds to  values obtained in the simulation by Jain et al. (1995).
Transition points were obtained by finding the intersection of straight lines 
which we fit to represent various regimes. }
\end{figure}

\section{Unified analysis for different regimes}

In the last section, we discussed the intermediate and nonlinear regimes separately. It is, however, possible to discuss the two rgimes together by using a simple approximation. We shall discuss this approach in this section. The results of last section can be obtained as a special case of this approach.
To do this we begin with equation (\ref {three}) written  as
\begin{equation}
{\partial D \over \partial  A} - h{\partial D \over \partial  X} = 3h
\label{twentynine}
\end{equation}
where we have introduced the following new variables

\begin{equation}
 D = \ln[(1 + \bar \xi_2(x,a)], ~~~~ A = \ln a, ~~~~  X = \ln x
\end{equation}

\n
and written the pair velocity as  $ v = -h \dot a x $.
Simulations indicate that  $h \approx 2$  in intermediate regime and and $h \approx 1$ in
 nonlinear regime which we have used for results obtained in earlier sections. Here we will try to get an unified picture covering both the regimes.
We shall now assume that we can treat $h$ as approximately constant while integrating this equation. In that case, the general solution is

\begin{equation}
1 + \bar \xi_2  = a^{3 h} F ( a^h x )\label{thirtythree}
\end{equation}

\n
where  $F$ is an  arbitrary function to be determined  by initial condition.
If the linear $\bar \xi_L$ is a power law, we know tht the true $\bar\xi_2(a, x)$ can only depend on the variable $q \equiv xa^{-2/(n+3)}$ which is possible only if $F$ is a powerlaw. So we {\it must} have 
\begin{equation}
\xi_2(x,a) \propto a^{3h} ({ax^h})^{-\gamma}
\end{equation}

\n
The index $\gamma$ can be determined by matching the above expression with linear two point correlation function  at a scale $x_c = a^{2/(n + 3)}$ which is going  
nonlinear. We then get

\begin{equation}
\gamma = {{ 3h(n+3) }\over{h(n+3) + 2} }
\end{equation}

\n
Now it is possible to write the two point correlation function 
as 

\begin{equation}
\bar \xi_2(x,a) \propto a^{6h/[ 2 + h(n+3)]} x^{-3h(n+3)/[2 + h(n+3)]}
\end{equation}
Earlier results of 
intermediate and highly non-linear regime can be recovered by taking
$ h = 2 $ and $ h = 1 $ respectively.

It is actually possible to relax the power law requirement and still obtain a general result.
From the characteristics of 
equation (\ref {twentynine}) we can show that $\bar\xi_2 (x, a)$ can be expressed as a function of $\bar\xi_L (l, a)$ where     $l^3 \equiv x^3 [1 + \bar\xi_2 (x, a)]$. That is 
\begin{equation}
\bar \xi_2(x,a) = U[\bar \xi_L(l,a)]
\end{equation}

\n
where $ U$ is some function. Combining with (\ref {thirtythree}) we have  $a^{3h}F(a^hx) = U[\bar \xi_L(l,a)]$ or equivalently
$a^{3h}F(r) = U[a^2 Q(r)]$ where we have used the fact that we can write
$l^3 \cong x^3\bar \xi_2 = r^3F(r)$ with $ r = a^h x$.
But the above expression can be valid for arbitrary $a$ at fixed $r$, only if

\begin{equation}
U(z) \propto z ^{3h/2}
\end{equation}

\n
So, in terms of correlations functions, we must have   
\begin{equation}
\bar \xi_2(x,a) \propto [ \bar \xi_L(l,a) ]^{3h/2}
\end{equation}
This generalises the relations $\bar\xi_2 \propto \bar \xi^3_L$, $\bar\xi_2 \propto \bar\xi_L^{3/2}$ we used for quasilinear and nonlinear regimes in 
the last section. 

To perform the averages over regions with different peak heights, we only
have to do the rescaling $\bar\xi_L \to \nu^2 \bar\xi_L$ and note that 
$\bar\xi_2 \propto (x_i/x)^3 \propto \bar\xi_L^{3h/2} (x_i)$ implies 
$x\propto x_i \bar \xi_L^{h/2} (x_i)$. Simple algebra then gives

\begin{equation}
\xi_2(x,a) \propto {\langle x_i^3 \rangle \over x^3} \propto 
\langle \nu^{6h/(2 + h(n+3))} \rangle x^{-3h(n+3)/( 2 + h(n+3))}
\end{equation}
This generalises the relation  (\ref {xitwo}) and (\ref {xibartwo}) of the last section. Assuming that $\nu$ is a Gaussian variable and performing the average, we get  upto a normalization,

\begin{equation}
\xi_2(x,a) \propto 2^{\alpha/2} \Gamma \big ( {{\alpha + 1} \over 2} )
x^{ - 3h(n+3) / {(2 + h(n+3))}}
\end{equation}
where $\alpha = 6h/[2+h(n+3)]$. Normalizing the expression properly, we can
write the final result as

\begin{equation}
\xi_2(a,x) =  {\cal A}(h,n) [ \xi_2(l, a)]^{3h/2}
\end{equation}

\n
where
\begin{equation}
{\cal A}(h,n) = \Big ( { 2 \over \lambda} \Big )^{3h \over 2} \Big[  {\Gamma \big( {\alpha +1
\over 2 }\big )\over 2 \sqrt \pi} \Big ]^{3h \over \alpha}
\end{equation}
To obtain the earlier results for intermediate regime we have to set
$h = 2$ which gives $ \alpha = 6/(n+4)$ and

\begin{equation}
A = \Big ( { 2  \over {\lambda} } \Big )^3 \Big[  {\Gamma \big( {\alpha +1
\over 2 }\big )\over 2 \sqrt \pi} \Big ]^{6 \over \alpha};
\end{equation}

\n similarly for the  nonlinear regime we have $h = 1$, 
$\alpha = 6/(n+5)$ and

\begin{equation}
B = \Big ( { 2  \over {\lambda} } \Big )^{3/2} \Big[  { \Gamma\big( {\alpha +1
\over 2 }\big )\over 2 \sqrt \pi} \Big ]^{3 \over \alpha}
\end{equation}

\n
These results match with earlier expressions. 

Using our ansatz in (\ref {xibarn}) it is possible to generalize the result for higher order correlation functions. We find

\begin{equation}
S_N(h,n) = \langle \nu^{\alpha(N-1)} \rangle / \langle \nu^\alpha \rangle^{N-1}
\end{equation}

\n which can be explicitely written as

\begin{equation}
S_N(h,n) = ( 4 \pi )^{(N - 2)/2} { \Gamma \big ( { \alpha ( N - 1 ) + 1 \over 2})
\over \Gamma \big ( { \alpha + 1 \over 2})}
\end{equation}

\n
This allows calculation of $S_N(\xi_2(x,a))$, given $h( \xi_2(x,a))$.

\section{Comparison with simulations}

\begin{figure*}
\protect\centerline{
\epsfysize = 4.05truein
\epsfbox[20 146 587 714]
{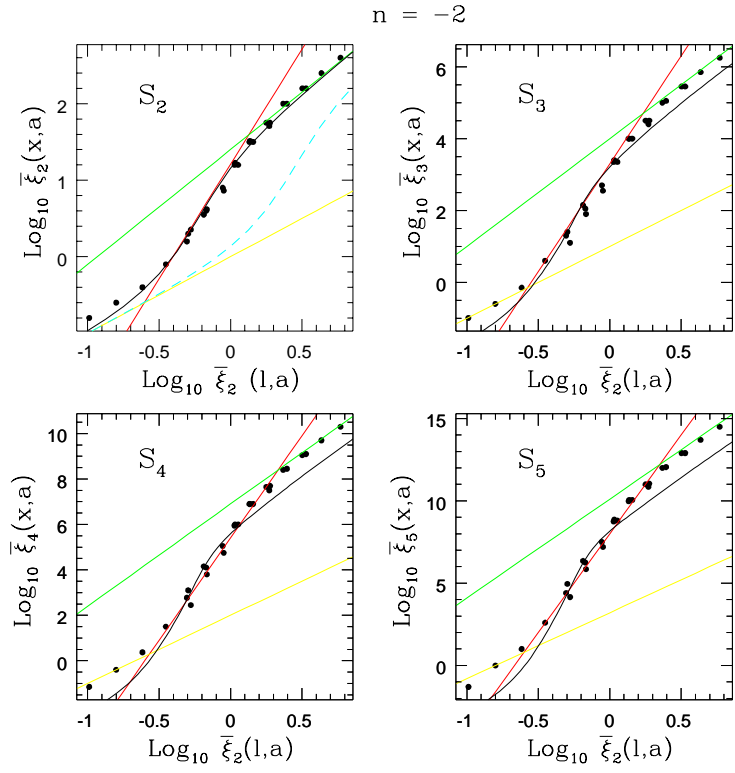}
}
\caption{ $\bar \xi_N(x,a)$ has been plotted against $\bar \xi_2(l,a)$ for $n = -2$ 
spectra. Solid curve in panel $S_2$ corresponds to fit by Jain et. al. ( 1995)
and dash curve for Hamilton et. al. ( 1991), in other panels solid curves correspond to prediction from our model. Straight lines in different
panel correspons to slopes $( N - 1)$, $3(N-1)$ and $3(N-1)/2$ 
for perturbative, intermediate and highly nonlinear regime respectively.
Dots represent N-Body simulation data from Colombi et. al. ( 1996 )  }
\end{figure*}

\begin{figure*}
\protect\centerline{
\epsfysize = 4.05truein
\epsfbox[20 146 587 714]
{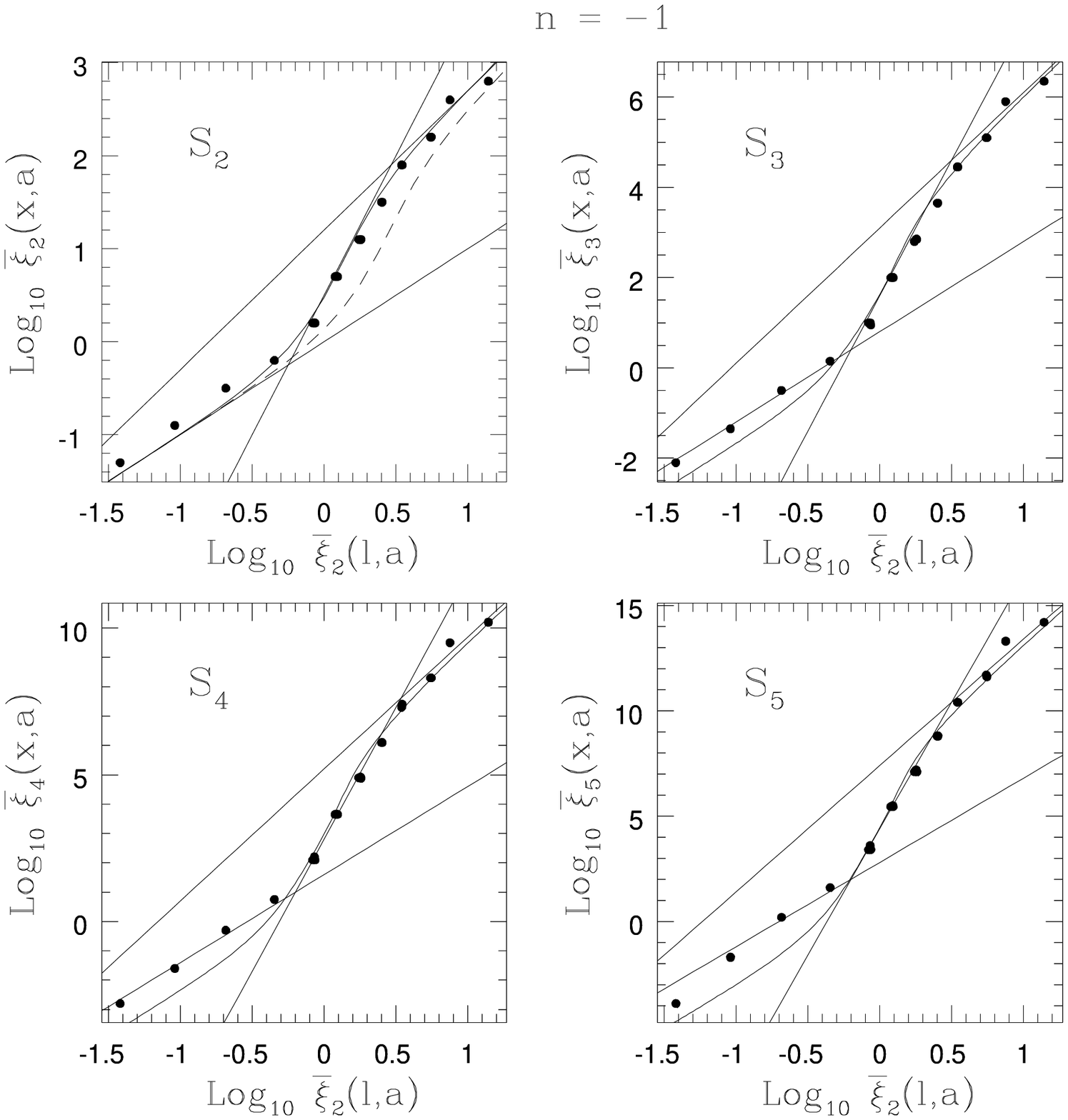}
}
\caption{ Same as figure 3 for $n = -1$ spectrum }
\end{figure*}

\begin{figure*}
\protect\centerline{
\epsfysize = 4.05truein
\epsfbox[20 146 587 714]
{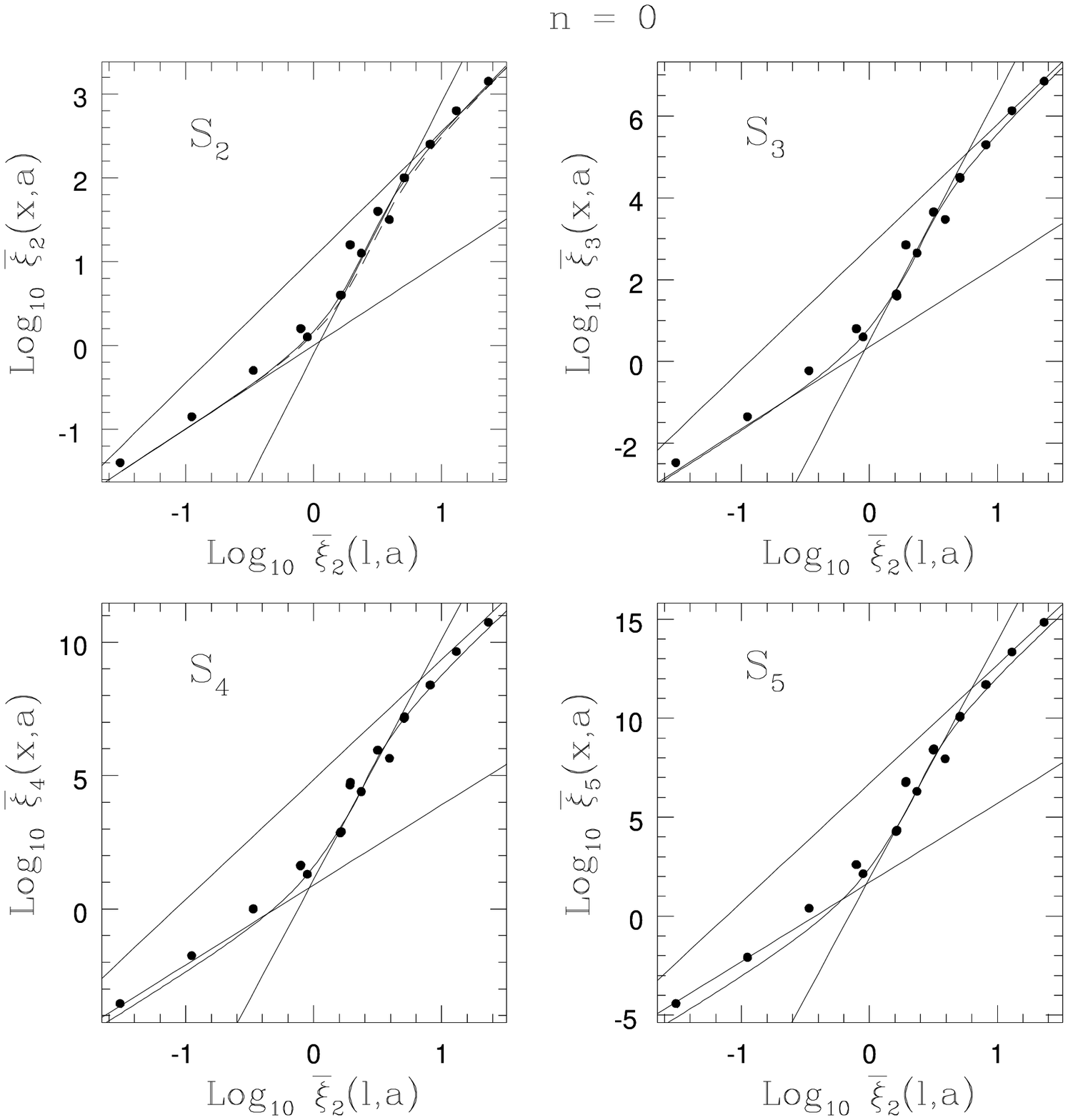}
}
\caption{ Same as figure 3 for $n = 0$ spectrum }
\end{figure*}

\begin{figure*}
\protect\centerline{
\epsfysize = 4.05truein
\epsfbox[20 146 587 714]
{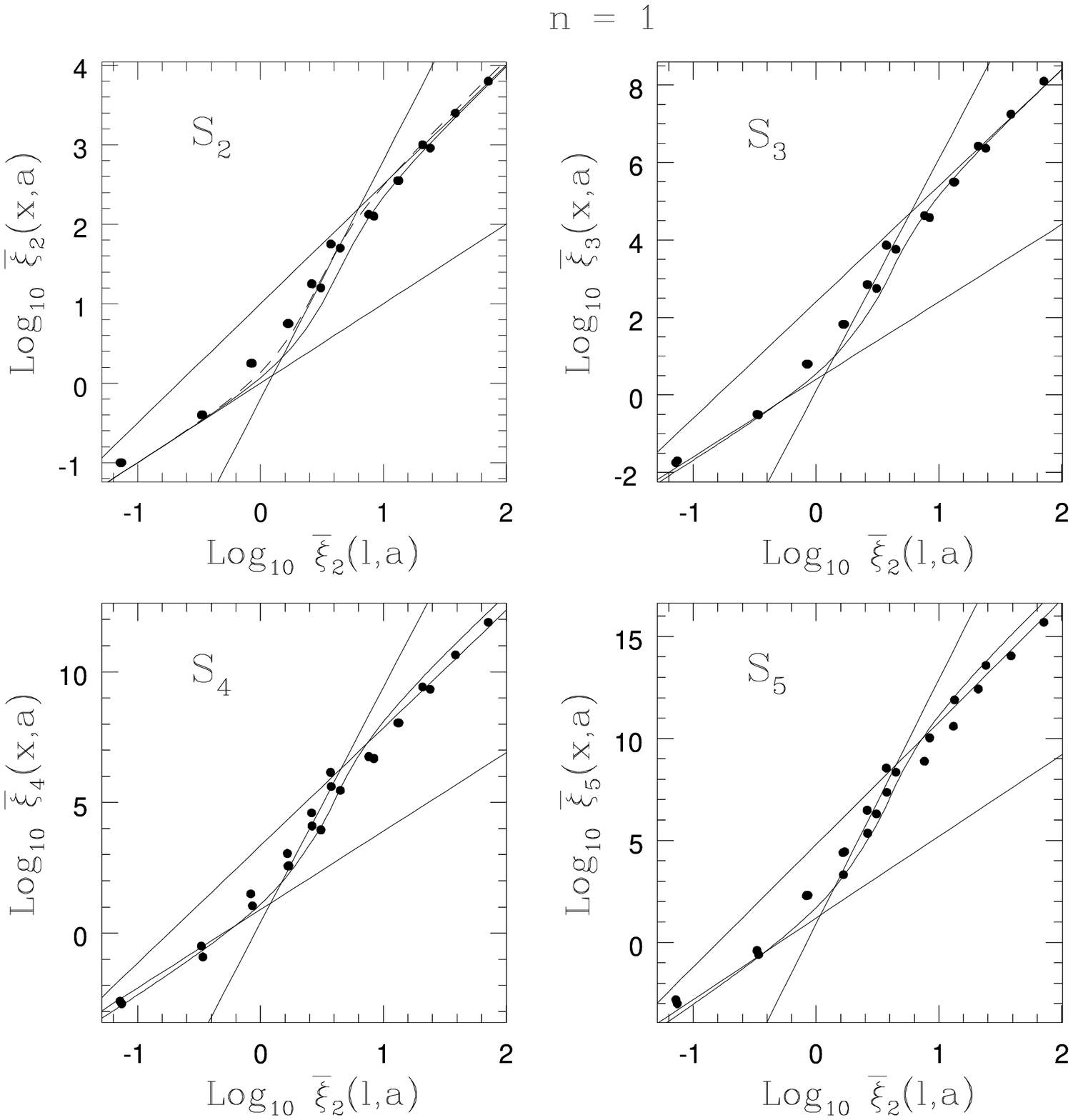}
}
\caption{ Same as figure 3 for $n = 1$ spectrum}
\end{figure*}

The results obtained in the previous section can be compared with the
simulations as regards two  essential aspects. 

First of all, the results show that the
spectral dependence of the scaling relations between the nonlinear and
linear correlation functions can arise due to averaging peaks of different heights. This, in turn, implies that the scales
at which the transition from  perturbative regime to intermediate regime, or from intermediate regime to nonlinear regime takes place depends on the power
spectrum index. By comparing the predicted values for these transitions with
the result of simulations, we can test the the validity of our averaging scheme
and the basic model for the two point correlation function. Secondly, we can compare the values of $\bar\xi_N$ obtained from the model with those of simulation
in both intermediate and nonlinear regimes. Since $\bar\xi_2$ is related to $\bar\xi_{2L}$ in a nonlocal manner, and our ansatz relates $\bar\xi_N$ to 
$\bar\xi_2$ (in  an indirect manner), we would expect some nonlocal relationship between $\bar\xi_N$ and $\bar\xi_2$.  This will test the validity of our
ansatz regarding  higher order correlation functions. Note that these two
comparisons test the two distinct generalisations of the work in
Padmanabhan (1996), introduced in this paper.  

The comparison of predicted values for the transition is shown  in figure 1 along with results of numerical simulation given
in Jain et al. (1995). We see that there is good agreement between theory
and simulations suggesting that (i) the basic picture for the evolution of
gravitational clustering, developed in Padmanabhan (1996) is correct and 
(ii) the spectrum depedence of the scaling relation can be understood
by averaging over the initial fluctuations. The analysis also shows that
 --- as $n_p$ changes from -2 to 0 --- the lower transition point varies
between $0.25$ and $1.0$ while the upper one varies between $2.0$ and $7.0$

We shall next turn to the comparison of $\bar\xi_N$  predicted by our
model with the results of numerical simulations, in order to test the
validity of our ansatz.
In doing so, one should be aware of several effects which could ``corrupt"
the values of $S_N$ parameters in the simulations, and take adequate
precautions to correct for them. At small  $\sigma$, i.e. 
in the perturbative regime, the main contribution to error comes from cosmic variance
i.e. due to presence of small number of large cells containing completely
independent samples. This error can be reduced only by increasing
the size of N-body computation box. On the other hand in the highly 
nonlinear regime one is restricted by resolution of the N-body simulation
for probing very small scale. Poisson shot noise also starts playing 
increasingly dominant role as soon as the average occupancy of cells 
becomes comparable to unity. The usual procedure used for computing $S_N$ parameters
is by taking moments of cell counts P(n) for different cell sizes. In general,
cell counts  show a power law behaviour in the highly 
non linear regime upto $n = n_c (= \bar n \bar\xi_2)$ followed by an exponential 
tail at $ n>n_c$ (Balian \& Schaeffer 89). In an ideal (infinite) catalogue this exponential tail will be extended to 
very small values of P(n); but, in practice, there is a sharp cutoff 
around $n = n_{max}$ which is the most dense cell present in the N-body 
catalogue. Higher moments of P(n) - and hence higher $S_N$ parameters - are
more sensitive to large n tail of P(n). So it is extremely important
to extend the the exponential tail to infinity and then normalise
the corrected P(n) again before calculating $S_N$ parameter. This way of 
correcting for measured $S_N$ has been extensively studied and expected to 
give correct - or at least, more reliable - result (Colombi et al. 92, 94, 95,
Lucchin et al. 94 ).
Recently Colombi et al. ( 1995 ) has done a careful analysis of high resolution n-body
data with large dynamic range correcting for all the errors mentioned above. 
Their study covers power law models $n_p$ = 1, 0, -1, -2. and they study 
first three non-trivial $S_N$ parameters i.e. $S_3$, $S_4$, and $S_5$. 

We have computed $\bar\xi_N$ for $N= 2,3,4,5$ for the power laws 
$n = -2, -1, 0, 1$ using the published data in Colombi etal. (1995) and 
our ansatz. The results are shown in figures 2 to 5. The 
theoretical prediction based on our ansatz is shown by solid, S-shaped line 
and the simulation data is shown by points. The straight lines have the 
asymptotic theoretical slopes.

It is clear from the graphs that our basic ansatz is an attempt in the 
right direction. The overall agreement between the theory and simulations is 
good especially when we consider the simplicity of our model. We will now 
comment on several details related to the comparison between theory and 
simulations.

Plots of $\bar \xi_N(x,a)$ vs $\bar \xi_2(x,a)$ shows that our basic claim regarding three
phases in graviational clustering seems to be correct and also one do get
a nonlocal scaling relation for N-point correlation function similar to
 scaling for two point correlation function as suggested from the characteristics. 
To some extent the scaling in higher order correlation function reflect
underlying scaling in two point correlation function; but it can also be
argued that if  $\bar \xi_N(x,a)$ can be expressed
as a smoothly varying function of $\bar \xi_2(x,a), ~~\bar \xi_N(x,a) = T_N( \bar \xi_2(x,a))$ 
then we can write $\bar \xi_N(x,a) = T_N( F_n(\bar \xi_2(l,a))) = G_{n,N}(\bar \xi_2(l,a))$
where $G_{n,N}= T_N*F_n$ and thus define a scaling relation between $\bar \xi_N(x,a)$
and $\bar \xi_2(l,a)$.

The relation between $\bar \xi_2(a,x)$ and $\bar\xi_2(l,a)$ shows good agreement with the analytical fit suggested
by Jain et. al. (1995)
The scatter in the data increases with $n$, which can be understood in following manner. We use $x$ and $\bar \xi_2(x,a)$ to recover the lagrangian radius 
$l = x( 1 + \bar \xi_2(x,a))^{1/3}$ which was then used to get the linearly extraploated
$\bar \xi_2(l,a) = \sigma_0^2a^2 l^{-(n+3)}$. Error in  estimation of $x$ or $\bar \xi_2(x,a)$ from the published data of Colombi et al. (1995) gets reflected finaly in error of $\bar \xi_2(l,a)$. We can relate
fractional error $\Delta_{\bar \xi_2(l,a)}$ with fractional error
$\Delta_l$ by 
$|\Delta_{\bar \xi_2(l,a)}| = (n+3) | \Delta_l |$
Which shows that for same $\Delta_l$, $\Delta_{\bar \xi_2(l,a)}$ increase with
n. This explains (partly) why we get more scatter for $n = 1$ spectra compared to
$n = - 2$ spectra.

\section*{Acknowledgment}
It is pleasure for D.M. to acknowledge his thesis supervisor, Varun Sahni,
for constant encouragement and active support during the course of work.
D. M. thanks Francis Bernardeau and Richard Schaeffer for many useful discussions and warm hospitality during his stay at CEA ( Saclay ). 
 ~  D. M. was financially supported by the Council of
Scientific and Industrial Research, India, under its SRF scheme.

\section*{ Appendix - A}

Given the $S_N$ parameters, one can compute the void probability function
which is of some theoretical and practical importance. In scaling models (which assume $S_N$ parameters are constant over some length scale) 
the void probability function can be written as ( White 1979 ) 

\begin{equation}
P_0 = \exp ( -\phi(n_c)/ \bar \xi_2 )
\end{equation}
where $\phi$ is generating function for
$S_N$ parameters and is defined as 

\begin{equation}
\phi(n_c) = -\sum_{p=1}^{\infty}{ S_p \over p!} (-n_c)^p
\end{equation}
where $n_c = \bar n \bar \xi_2$ is a scaling variable. 
Substituting our expression for $S_N$ in the above equation we get

\begin{equation}
\phi(n_c) = n_c- \sum_{p=2}^{\infty} (4\pi)^{(p-2)/2}{ \Gamma(((p-1)s + 1)/2)
\over \Gamma( (s + 1)/2))^{p - 1} p!} (-n_c)^p
\end{equation}
where $s = z$ for intermediate regime and  $s = y$ for nonlinear regime.
Using definition of gamma function it is easy to write down the sum as 
\begin{eqnarray}
\phi(n_c) &=& n_c - { \Gamma [ ( s + 1)/2] \over 4 \pi } \times \nonumber \\
 \sum_{p=2}^{\infty} && \int_0^{\infty} { dt  t^{(p-1)s/2}\exp(-t)
\over {\sqrt t}~p!} \bigg ( - {{ \sqrt {4 \pi} }~n_c~ \over \Gamma [(s + 1)/2]}
\bigg )^p 
\end{eqnarray}
Interchanging the sum and the integral 
we get
\begin{eqnarray}
&\phi(n_c)& = n_c - { \Gamma [ ( s + 1)/2] \over 4 \pi }  \times \nonumber \\
 &&\int_0^{\infty}  dt  t^{-(s+1)/2}e^{-t}(\exp(-f t^{s/2}) + f t^{s/2}-1)
\end{eqnarray}
where
\begin{equation}
f= \bigg (  {{ \sqrt {4 \pi} }~n_c~ \over \Gamma [(s + 1)/2]}
\bigg )
\end{equation} 
and taking the limit $n_c \rightarrow \infty$
shows that $\phi(n_c)\approx  n_c/2$ for large $n_c$. 
In the scaling model proposed by  Balian and Schaeffer, $\phi$ is expected
to scale as $n_c^{1-\omega}$ asymptotically; hence our ansatz leads to the
$\omega = 0$ model. (Note that 
other extreme case is $\omega = 1$ which is the negative binomial model proposed
earlier). The counts-in-cell $P(n)$ is proportional to $n^{\omega -2}$ showing
that, in our model,  $P(n) \propto n^{-2}$. 


\end{document}